\documentclass[aps,prc,reprint,secnumarabic,showpacs,floatfix,amsmath,amssymb,superscriptaddress,nofootinbib]{revtex4-1}  %for review and submission

\usepackage{graphicx,ragged2e}                    % needed for figures
\usepackage{dcolumn,multirow,floatrow}            % needed for some tables
\usepackage{bm}                                   % for math
\usepackage{amssymb,amsmath}                % for math
\usepackage{mathtools,array,subfigure}		  % some more math adjustments		
\DeclarePairedDelimiter{\abs}{\lvert}{\rvert}     % to make '\abs' works	
\newcolumntype{C}[1]{>{\centering\let\newline\\\arraybackslash\hspace{0pt}}m{#1}}  % column centering with fixed column-width
\usepackage{hyperref}                             % for highlighting refs, plots and links
\hypersetup{
    colorlinks=true,
    linkcolor=red,
    filecolor=blue,      
    urlcolor=magenta,
    citecolor=blue
    }
\makeatletter
\let\cat@comma@active\@empty
\makeatother

\begin{document}

\title{Exclusive $\pi^{+}$ electroproduction off the proton from low to high $-t$}

%%%% authors list from a separate .tex file (issue with affiliation{} and math mode)
%\input authors_list.tex
%===========================
%%%%%%%%%%%%%%%%%%%%First Author%%%%%%%%%%%%%%%%%%%%%%%%%%%%%%%%%%%%%
\author{S.~Basnet \thanks{}}
\email[Email:~]{basnet2s@uregina.ca}
\affiliation{University~of~Regina,~Regina,~SK,~S4S~0A2,~Canada}
\affiliation{Universit\'e~catholique~de~Louvain,~Louvain-la-neuve~1348,~Belgium}

%%%%%%%%%%%%%%%%%%%%Primary Authors%%%%%%%%%%%%%%%%%%%%%%%%%%%%%%%%%%
\author{G.M.~Huber \thanks{}}
\email[Email:~]{huberg@uregina.ca}
\affiliation{University~of~Regina,~Regina,~SK,~S4S~0A2,~Canada}

\author{W.B.~Li}
\affiliation{University~of~Regina,~Regina,~SK,~S4S~0A2,~Canada}
\affiliation{College~of~William~and~Mary,~Williamsburg,~Virginia~23187,~USA}

\author{H.P.~Blok}
\affiliation{VU~University,~1081HV~Amsterdam,~The~Netherlands}
\affiliation{National~Institute~for~Subatomic~Physics~(Nikhef),~1009DB~Amsterdam,~The~Netherlands}

\author{D.~Gaskell}
\affiliation{Thomas~Jefferson~National~Accelerator~Facility,~Newport~News,~Virginia~23606,~USA}

\author{T.~Horn}
\affiliation{Thomas~Jefferson~National~Accelerator~Facility,~Newport~News,~Virginia~23606,~USA}
\affiliation{Catholic~University~of~America,~Washington,~DC~20064,~USA}

%%%%%%%%%%%%%%%%%%%%%%Secondary Authors%%%%%%%%%%%%%%%%%%%%%%%%%%%%%%%
\author{K.~Aniol}
\affiliation{California~State~University~Los~Angeles,~Los~Angeles,~California~90032,~USA}

\author{J.~Arrington}
\affiliation{Argonne~National~Laboratory,~Argonne,~Illinois~60439,~USA}

%\author{B.~Barrett}
%\affiliation{Saint Mary's University, Halifax, Nova Scotia $\:B3H\:3C3$, Canada}

\author{E.J.~Beise}
\affiliation{University~of~Maryland,~College~Park,~Maryland~20742,~USA}

\author{W.~Boeglin}
\affiliation{Florida~International~University,~University~Park,~Florida~33199,~USA}

\author{E.J.~Brash}
\affiliation{Christopher~Newport~University,~Newport~News,~Virginia~23606,~USA}

\author{H.~Breuer}
\affiliation{University~of~Maryland,~College~Park,~Maryland~20742,~USA}

\author{C.C.~Chang}
\affiliation{University~of~Maryland,~College~Park,~Maryland~20742,~USA}

\author{M.E.~Christy}
\affiliation{Hampton~University,~Hampton,~Virginia~23668,~USA}

\author{R.~Ent}
\affiliation{Thomas~Jefferson~National~Accelerator~Facility,~Newport~News,~Virginia~23606,~USA}

\author{E.~Gibson}
\affiliation{California~State~University,~Sacramento,~California~95819,~USA}

\author{R.J.~Holt}
\affiliation{Argonne~National~Laboratory,~Argonne,~Illinois~60439,~USA}

\author{S.~Jin}
\affiliation{Kyungook~National~University,~80~Daehakro,~Bukgu,~Daegu~41566,~Korea}

\author{M.K.~Jones}
\affiliation{Thomas~Jefferson~National~Accelerator~Facility,~Newport~News,~Virginia~23606,~USA}

\author{C.E.~Keppel}
\affiliation{Thomas~Jefferson~National~Accelerator~Facility,~Newport~News,~Virginia~23606,~USA}
\affiliation{Hampton~University,~Hampton,~Virginia~23668,~USA}

\author{W.~Kim}
\affiliation{Kyungook~National~University,~80~Daehakro,~Bukgu,~Daegu~41566,~Korea}

\author{P.M.~King}
\affiliation{University~of~Maryland,~College~Park,~Maryland~20742,~USA}

\author{V.~Kovaltchouk}
\affiliation{University~of~Regina,~Regina,~SK,~S4S~0A2,~Canada}

\author{J.~Liu}
\affiliation{University~of~Maryland,~College~Park,~Maryland~20742,~USA}

\author{G.J.~Lolos}
\affiliation{University~of~Regina,~Regina,~SK,~S4S~0A2,~Canada}

\author{D.J.~Mack}
\affiliation{Thomas~Jefferson~National~Accelerator~Facility,~Newport~News,~Virginia~23606,~USA}

\author{D.J.~Margaziotis}
\affiliation{California~State~University~Los~Angeles,~Los~Angeles,~California~90032,~USA}

\author{P.~Markowitz}
\affiliation{Florida~International~University,~University~Park,~Florida~33199,~USA}

\author{A.~Matsumura}
\affiliation{Tohuku~University,~Sendai,~Miyagi~Prefecture~980-8577,~Japan}

\author{D.~Meekins}
\affiliation{Thomas~Jefferson~National~Accelerator~Facility,~Newport~News,~Virginia~23606,~USA}

\author{T.~Miyoshi}
\affiliation{Tohuku~University,~Sendai,~Miyagi~Prefecture~980-8577,~Japan}

\author{H.~Mkrtchyan}
\affiliation{Yerevan~Physics~Institute,~375036~Yerevan,~Armenia}

\author{I.~Niculescu}
\affiliation{James~Madison~University,~Harrisonburg,~Virginia~22807,~USA}

\author{Y.~Okayasu}
\affiliation{Tohuku~University,~Sendai,~Miyagi~Prefecture~980-8577,~Japan}

\author{L.~Pentchev}
\affiliation{College~of~William~and~Mary,~Williamsburg,~Virginia~23187,~USA}

\author{C.~Perdrisat}
\affiliation{College~of~William~and~Mary,~Williamsburg,~Virginia~23187,~USA}

\author{D.~Potterveld}
\affiliation{Argonne~National~Laboratory,~Argonne,~Illinois~60439,~USA}

\author{V.~Punjabi}
\affiliation{Norfolk~State~University,~Norfolk,~Virginia~23504,~USA}

\author{P.~Reimer}
\affiliation{Argonne~National~Laboratory,~Argonne,~Illinois~60439,~USA}

\author{J.~Reinhold}
\affiliation{Florida~International~University,~University~Park,~Florida~33199,~USA}

\author{J.~Roche}
\affiliation{Thomas~Jefferson~National~Accelerator~Facility,~Newport~News,~Virginia~23606,~USA}

%\author{P.G.~Roos}
%\affiliation{University of Maryland, College Park, Maryland$\:20742$, USA}

\author{A.~Sarty}
\affiliation{Saint~Mary's~University,~Halifax,~Nova~Scotia~B3H~3C3,~Canada}

\author{G.R.~Smith}
\affiliation{Thomas~Jefferson~National~Accelerator~Facility,~Newport~News,~Virginia~23606,~USA}

\author{V.~Tadevosyan}
\affiliation{Norfolk~State~University,~Norfolk,~Virginia~23504,~USA}

\author{L.G.~Tang}
\affiliation{Thomas~Jefferson~National~Accelerator~Facility,~Newport~News,~Virginia~23606,~USA}
\affiliation{Hampton~University,~Hampton,~Virginia~23668,~USA}

\author{V.~Tvaskis}
\affiliation{VU~University,~1081HV~Amsterdam,~The~Netherlands}

%\author{S.~Vidakovic}
%\affiliation{University of Regina, Regina, SK $\:S4S\:0A2$, Canada}

\author{J.~Volmer}
\affiliation{VU~University,~1081HV~Amsterdam,~The~Netherlands}
\affiliation{DESY,~22607~Hamburg,~Germany}

\author{W.~Vulcan}
\affiliation{Thomas~Jefferson~National~Accelerator~Facility,~Newport~News,~Virginia~23606,~USA}

\author{G.~Warren}
\affiliation{Thomas~Jefferson~National~Accelerator~Facility,~Newport~News,~Virginia~23606,~USA}

\author{S.A.~Wood}
\affiliation{Thomas~Jefferson~National~Accelerator~Facility,~Newport~News,~Virginia~23606,~USA}

\author{C.~Xu}
\affiliation{University~of~Regina,~Regina,~SK,~S4S~0A2,~Canada}

\author{X.~Zheng}
\affiliation{Argonne~National~Laboratory,~Argonne,~Illinois~60439,~USA}

\collaboration{The Jefferson Lab F$_{\pi}-$2 Collaboration}
\noaffiliation
%===========================

%%%% date
\date{\today}

%%%%% abstract (in new format)
\begin{abstract}

\begin{description} 
  
\item[Background] Measurements of exclusive meson production are a useful tool in the study of hadronic structure. In particular, one can discern the relevant degrees of freedom at different distance scales through these studies.    

\item[Purpose] To study the transition between non-perturbative and perturbative Quantum Chromodyanmics as the square of four momentum transfer to the struck proton, $-t$, is increased.

\item[Method] Cross sections for the $^{1}$H$(e,e'\pi^+)n$ reaction were measured over the $-t$ range of 0.272 to  2.127 GeV$^{2}$ with limited azimuthal coverage at fixed beam energy of 4.709 GeV, $Q^{2}$ of 2.4 GeV$^{2}$ and $W$ of 2.0 GeV at the Thomas Jefferson National Accelerator Facility (JLab) Hall C.

\item[Results] The $-t$ dependence of the measured $\pi^{+}$ electroproduction cross section generally agrees with prior data from JLab Halls B and C. The data are consistent with a Regge amplitude based theoretical model, but show poor agreement with a Generalized Parton Distribution (GPD) based model.

\item[Conclusion] The agreement of cross sections with prior data implies small contribution from the interference terms, and the confirmation of the change in $t$-slopes between the low and high $-t$ regions previously observed in photoproduction indicates the changing nature of the electroproduction reaction in our kinematic regime.

\end{description}

\end{abstract}

%%%% pac numbers

\maketitle

%%%% I. INTRODUCTION
\section{Introduction}
\label{i}

A central topic in contemporary intermediate-energy subatomic physics is the description of hadronic matter in terms of the partonic constituents (quarks, $q$, and gluons, $g$) of Quantum Chromodynamics (QCD). In particular, the interface between hadronic and partonic descriptions of the strong interaction is of primary interest since the confinement of quarks and gluons into hadrons ($qqq$, or $q\overline{q}$ objects that interact strongly) is yet to be explained by QCD in detail. Thus, one has to rely on experimental studies of hadronic reactions probing the transition region to better understand this QCD interface. The exclusive electroproduction of a meson from a nucleon, $\gamma^{*}N \to N'M$, offers an excellent way to perform such studies.    

%%%% Figure #1 (Feynman Toy Diagrams)
\begin{figure}[h!]
\includegraphics[width=\linewidth]{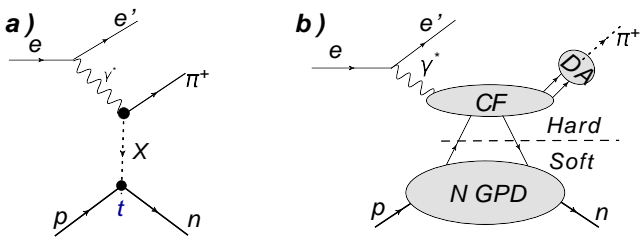}
	\caption{Exclusive $\pi^{+}$ electroproduction ($e + p \to e' + \pi^{+} + n$). In panel a), the $-t$ channel meson exchange Feynman-type diagram is shown, where $X\equiv$ exchange of Regge trajectories up to a cutoff scale, $\Lambda$. The so-called ``handbag diagram" is shown in panel b). The soft non-perturbative physics is contained in the GPD, whereas the hard scattering process, represented by collinear factorization (CF) and distribution amplitude (DA) in the figure, above the dotted line is calculable using pQCD. \iffalse Also, $W$ is the invariant mass of the system, whereas $t$ is the square of four momentum transfer to the nucleon. \fi}
\label{fig:feynman}
\end{figure}
%%%%

In particular, exclusive $\pi^{+}$ electroproduction off the proton provides
two ways to vary the interaction scale to study the interface between soft and
hard physics. Either the virtuality of the incoming photon, $Q^2 = -(p_e -
p^{'}_{e})^{2}$, effectively representing the transverse wavelength of the
photon probe ($\lambda \sim 1/Q$), or the square of four momentum transfer to
the nucleon
\footnote{For exclusive electroproduction reactions, four momentum transfer
  squared, $t$, is always negative and thus, the positive quantity, $-t$, will
  be used throughout this paper.}, 
$-t = (p_N -p^{'}_{N})^{2}$, representing the impact parameter ($b \sim
1/\sqrt{-t}$), can be varied independently. The invariant mass of the system is
given by $W = \sqrt{s}$, where the Mandelstam variable $s = (p_{e} - p^{'}_{e} +
p_{N})^{2}$. Here, $p_{e}$ and $p^{'}_{e}$ are the four-momenta of the initial
and scattered electron respectively, while $p_N$ and $p^{'}_{N}$ are the
initial and recoiled nucleon four-momenta respectively.

In the low $-t$ region ($\lesssim 0.9$ GeV$^{2}$), a description of hadronic degrees of freedom in terms of effective hadronic Lagrangians is valid. The effective theories take hadrons as the elementary particles, whose interactions are described by the exchange of mesons, as shown in Figure \ref{fig:feynman}(a). The virtual photon, $\gamma^{*}$, in this regime, behaves as a beam of vector mesons which passes far away from the nucleon target, i.e. large impact parameter ($b$), and the exchanged partons have enough time to hadronize into various mesons whose exchange primarily drives the cross section \cite{laget}. At higher $-t$ ($\gtrsim 0.9$ GeV$^{2}$), the impact parameter is small enough to force the partons to exchange a minimum number of gluons between the meson and the nucleon target before they recombine into the final particles. Hard-scattering processes such as these are at the origin of various factorization and scaling rules \cite{laget}. One such factorization is described by the so-called ``handbag diagram" (Figure \ref{fig:feynman}(b)), in which the complex quark and gluon non-perturbative structure of the nucleon is described by Generalized Parton Distributions (GPDs), while the hard process is factorized and calculable using perturbative QCD (pQCD). In this work, $Q^{2}$ was kept at a moderate nominal value (2.50 GeV$^{2}$), varying  $-t$ from near zero to 2.1 GeV$^{2}$, with the main aim to study the $-t$ dependence of the exclusive $\pi^{+}$ electroproduction cross sections. 

%%%%% Figure #2 (Kinematics)
\begin{figure*}[htpb]
	\mbox{
    \subfigure[\:\:$W-Q^{2}$ coverage at two  different $-t$ settings. The red data points represent $-t=0.272$ GeV$^{2}$, whereas the blue represent $-t=2.127$ GeV$^{2}$ setting.]{
     \includegraphics[width=0.30\linewidth]{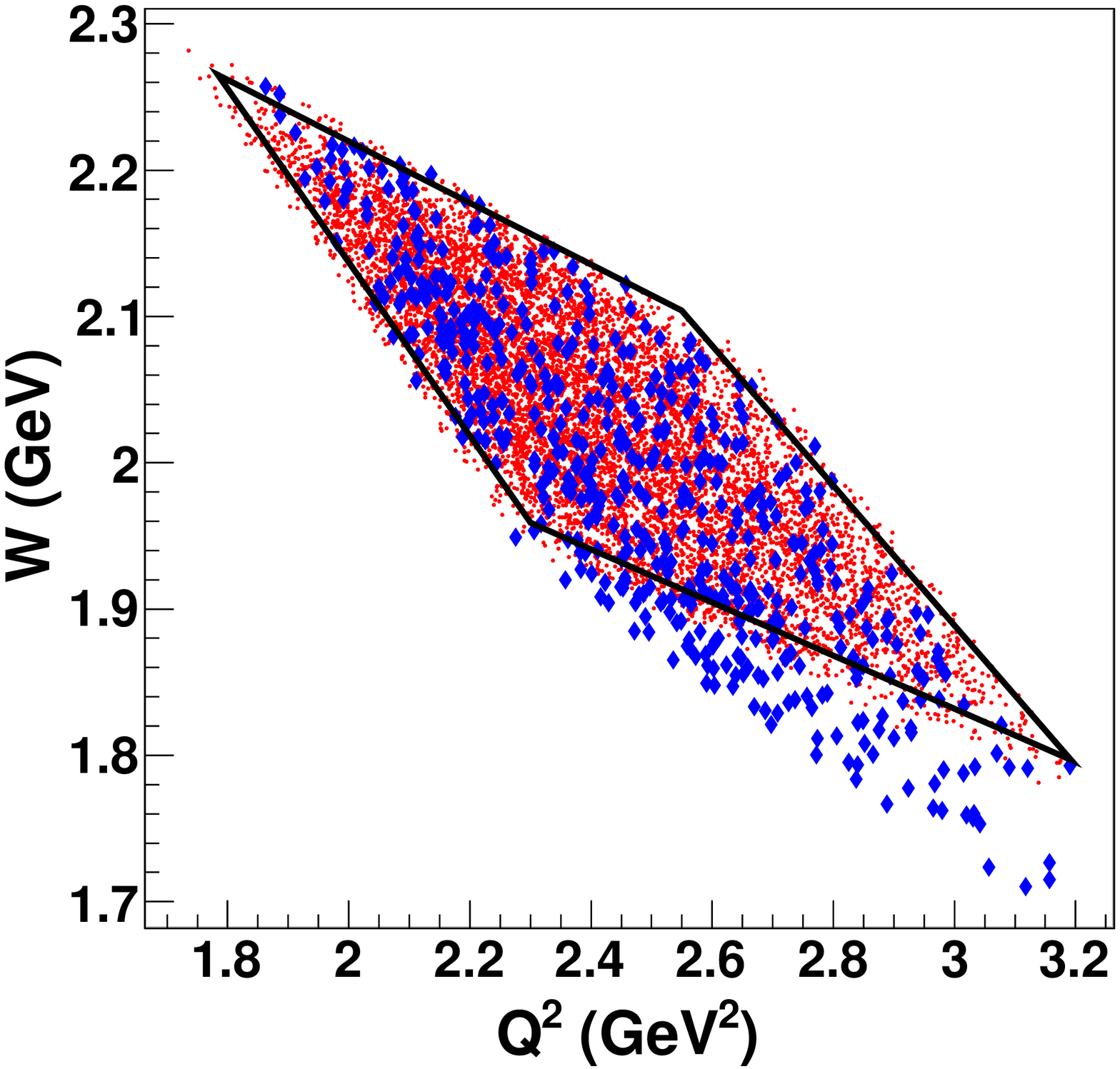}
      \label{kin1}
     	}
      
      \subfigure[\:\:$x_{B}-Q^{2}$ phase space at $-t=0.272$ GeV$^{2}$.]{
        \includegraphics[width=0.30\linewidth]{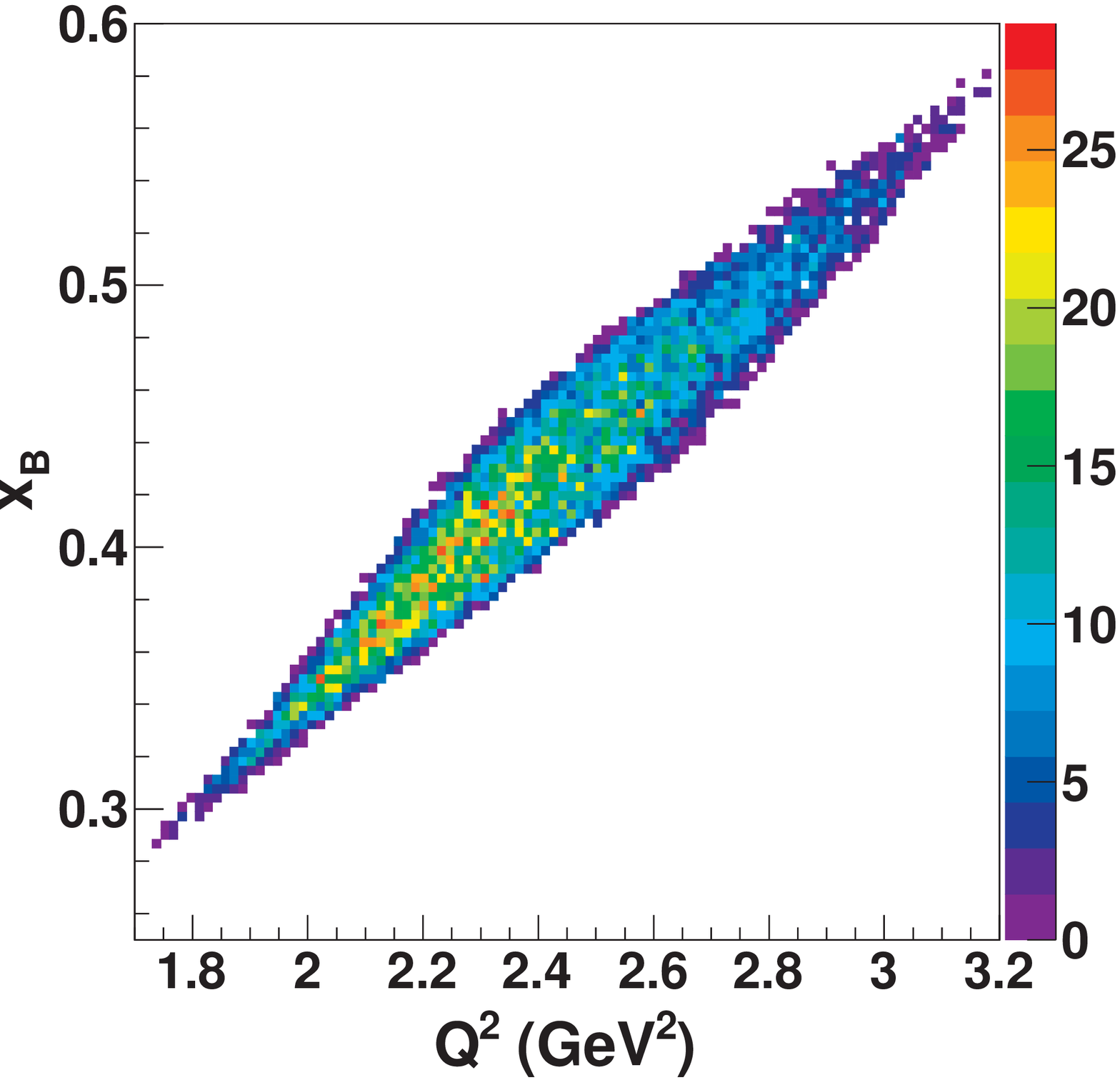}
        \label{kin2}
		  }
      \subfigure[\:\:$\abs{t}-\phi_{\pi}$ phase space coverage of the experiment represented in a polar plot. $-t$ is  plotted as the radial  component and $\phi_{\pi}$ as  polar component that progresses counter-clockwise, with $\phi_{\pi}$ = $0$ rad at the right.]{
        \includegraphics[width=0.34\linewidth]{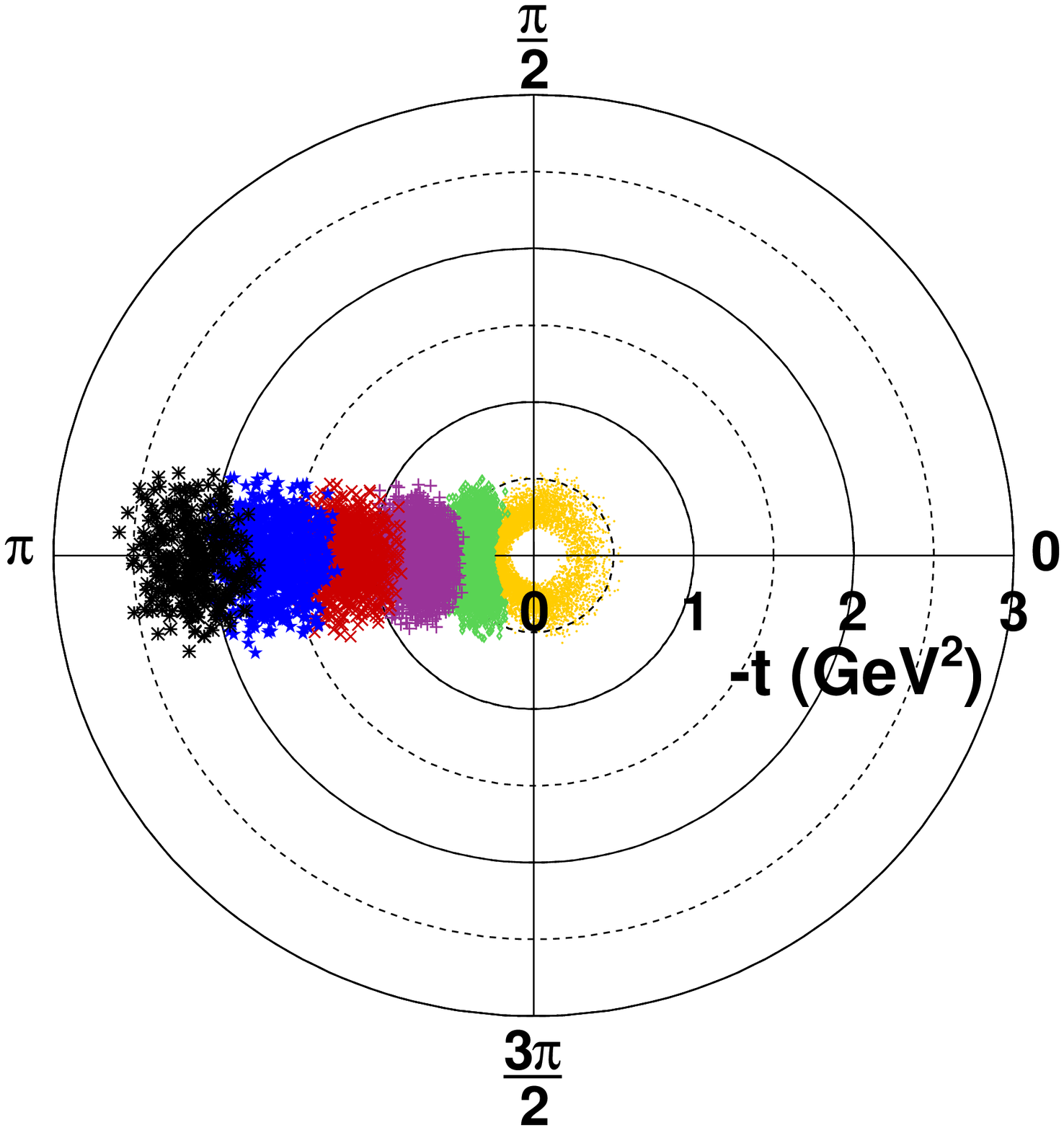}
        \label{kin3}
      }
	}
\caption{(Color Online) Kinematics of this measurement. See text for more details.}
%\caption{(Color Online) Kinematics of this measurement.  See text for more details.\\
%(a) $W-Q^{2}$ coverage at two  different $-t$ settings. The red data points
%represent $-t=0.272$ GeV$^{2}$, whereas the blue represent $-t=2.127$ GeV$^{2}$
%setting.
%(b) $x_{B}-Q^{2}$ phase space at $-t=0.272$ GeV$^{2}$.
%(c) $\abs{t}-\phi_{\pi}$ phase space coverage of the experiment represented in a
%polar plot. $-t$ is  plotted as the radial  component and $\phi_{\pi}$ as
%polar component that progresses counter-clockwise, with $\phi_{\pi}$ = $0$ rad
%at the right.}
\label{fig:kin}
\end{figure*}
%%%%

%%%% II. EXPERIMENTAL SETUP
\section{Experimental Setup}
\label{ii}

The data for exclusive $\pi^{+}$ electroproduction were acquired at the Thomas Jefferson National Accelerator Facility (JLab) Hall  C as a part of experiment E01-004, F$_{\pi}$-2 \cite{fpi2_th,q2_scale,fpi2_hb,fpi2_gh}. During the measurement, the unpolarized electron beam from the Continuous Electron Beam Accelerator Facility (CEBAF) of JLab, at fixed beam  energy of 4.079 GeV and beam current of 75 $\mu$A, was  incident on a 4 cm long liquid hydrogen (LH$_{2}$) target. Using two moderate  acceptance, magnetic focusing spectrometers, data for $^{1}$H$(e,e'\pi^{+})n$ were taken at a  central value of the virtuality of the incoming photon $Q^{2}$  of 2.5 GeV$^{2}$ and a central value of  the invariant mass $W$ of 2.0 GeV. The electrons  were detected in the Short Orbit Spectrometer (SOS), while the coincident electroproduced $\pi^{+}$ were detected in the High Momentum Spectrometer (HMS). The  measurement covers the $-t$ range from 0.272 to 2.127 GeV$^{2}$ at $x_{B}=0.44$ and $\varepsilon=0.56$. Here, $x_{B}$ is the fraction of the three momentum carried by the struck parton in the Breit frame and $\varepsilon$ is the longitudinal polarization of the virtual photon, given by Equation \ref{eqn:c}. Representative examples of $Q^{2}$, $W$, and $x_{B}$ coverage for the experiment are provided in Figures \ref{kin1},\ref{kin2}. To ensure that the acceptance weighted averages $\overline{Q^{2}}$ and $\overline{W}$ are the same  throughout all $-t$ settings, the so-called ``diamond cut" was applied, as shown in Figure \ref{kin1}. The available $x_{B}-Q^{2}$ phase space for $-t=0.272$ GeV$^{2}$ setting is shown in Figure \ref{kin2}. This particular setting was chosen because of the larger statistics compared to other settings. However, all $-t$ settings cover the same phase space.    

%%%% Table I (Kinematics + Results)
\begin{table*}[tbp]
\caption[Central kinematic settings  used for the exclusive $\pi^{+}$ electroproduction for this measurement.] {Central kinematics (four momentum transfer to the nucleon, $-t$, central hadron arm momentum, $P_{\pi}$, and scattering angle of the hadron arm, $\theta_{\pi}$)  of this exclusive $\pi^{+}$ electroproduction study. The weighted averages, $\overline{Q^{2}}$ and $\overline{W}$, of the data are also listed, along with the unpolarized cross section results (in $\mu$b/GeV$^{2}$) given in the last column. Two uncertainties for the cross section results are provided, with the first being the combination of statistical and uncorrelated systematic uncertainties added in quadrature, while the second one represents the correlated (scale) uncertainties.}
\label{table:kinematics}
\begin{tabular}{ccccc|C{3.2cm}}   
      $\abs{t}$	& $\overline{Q^{2}}$    & $\overline{W}$       & $P_{\pi}$	& $\theta_{\pi}$	  & $\frac{d^{2}\sigma} {dt \: d\phi_{\pi}} \rvert_{\phi_{\pi}=\pi}$ 			\\
      (GeV$^2$)	& (GeV$^2$) & (GeV)     & (GeV/$c$)	& (deg) 		          & ($\mu$b/GeV$^{2}$)   		\\ \hline 
      $0.272$		& $2.402$   & $2.039$   & $2.845$ 	& $15.68$   		      & $0.367$ $\pm$ $0.030, 0.013$	\\
      $0.378$		& $2.427$   & $2.029$   & $2.788$	  & $20.32$   		      & $0.288$ $\pm$ $0.051, 0.010$	\\
      $0.688$		& $2.449$   & $2.018$   & $2.622$	  & $25.15$   		      & $0.164$ $\pm$ $0.034, 0.006$	\\
      $1.145$		& $2.427$   & $2.029$   & $2.378$	  & $30.07$   		      & $0.096$ $\pm$ $0.006, 0.003$	\\
      $1.608$		& $2.433$   & $2.020$   & $2.131$	  & $34.50$   		      & $0.054$ $\pm$ $0.002, 0.002$	\\		              
      $2.127$		& $2.423$   & $2.026$   & $1.853$	  & $39.50$   		      & $0.032$ $\pm$ $0.002, 0.001$	\\
\end{tabular}
\end{table*}
%%%%

In order to study the $t$-dependence of the exclusive  pion electroproduction cross section, the central momentum, $P_{\pi}$, of the pion arm was varied from $2.845$  GeV/$c$ at the lowest $-t$ setting of $0.272$ GeV$^2$, to $1.853$ GeV/$c$ at the highest $-t$ setting  of 2.127 GeV$^2$. This was done in concert with the variation of the scattering angle, $\theta_{\pi}$, of the pion arm, from $15.68^{\circ}$ at the near-parallel ($\theta_{\pi q} \approx 0^{\circ}$, where $\theta_{\pi q}$, in the lab frame, is the angle between emitted pion and $q$ vector defined by the SOS)  kinematics, to $39.50^{\circ}$ at  the highest $-t$  setting, as shown  in Table \ref{table:kinematics}. The average $\overline{Q^{2}}$ and $\overline{W}$ for each $-t$ setting are also listed in the table. The  acceptances of the  two spectrometers at non-parallel kinematics do not  provide full coverage in $\phi_{\pi}$. The complete $\abs{t}$-$\phi_{\pi}$ coverage of our data is illustrated in Figure \ref{kin3}. They are centered around $\phi_{\pi} = \pi$, except at the lowest $-t$ setting.

In the experiment, electron identification was  done using a combination of gas Cherenkov detector and lead-glass calorimeter in the  SOS. $\pi^{+}$ identification in the HMS was largely done using  time of flight between two scintillating hodoscope arrays. In addition, an aerogel Cherenkov detector was used to further reject  proton events. Any remaining contamination by real  electron-proton coincidences was removed by a single  beam-burst cut on $e$-$\pi^{+}$ coincidence time. A more detailed description of both spectrometers, along with the respective detector stacks, can be found in Ref. \cite{fpi2_hb}.

%%%% III. DETERMINATION OF CROSS SECTION
\section{Determination of the cross section} 
\label{iii}

\begin{figure}[b!]
\includegraphics[width=\linewidth]{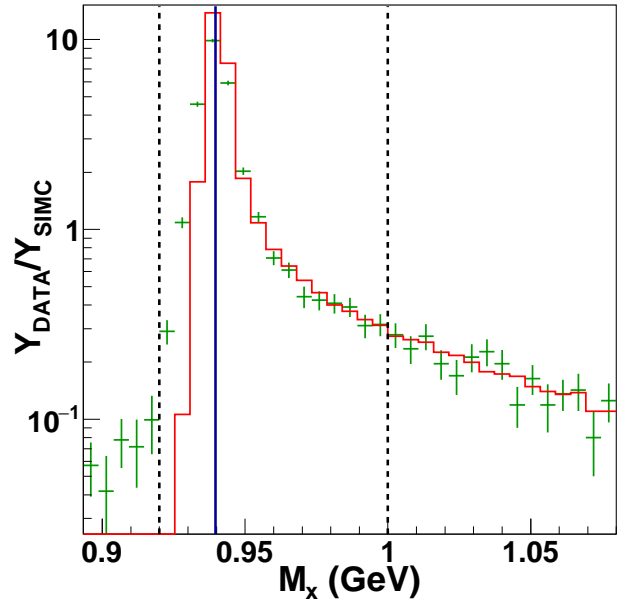}
  \caption{ (Color Online) Representative missing mass (M$_{x}$ in GeV) distribution at $-t$=0.378 GeV$^{2}$ setting. The data along with their errors are represented with green crosses and they agrees quite well with SIMC simulated result for the same setting represented by red histogram. The solid blue line is the mass of neutron \cite{pdg}, while the black dotted lines at $M_{x}=$0.92 and 1.00 GeV respectively represent the missing mass cut used.}
\label{fig:mx}
\end{figure}

The raw data  collected by the data acquisition  system were processed using  the standard Hall C analysis  software (ENGINE), which decodes the  raw data into physical  quantities on an event-by-event basis  in order to perform the necessary data  analysis. Some of the major components  of the analysis includes identification of good  events, spectrometer acceptance reconstruction, background subtraction  (from random coincidences and target cell), tracking  and particle identification.  These are discussed  extensively in Refs.  \cite{fpi2_th,fpi2_hb}. The relevant electroproduction  kinematic variables, such  as $Q^{2}$, $W$,  $-t$, were reconstructed  using the spectrometer quantities. Using energy and momentum conservation, the  exclusive $n\pi^{+}$ final state was reconstructed and the appropriate events were selected using a  cut on the missing mass ($M_{x}$) for the reaction (Figure \ref{fig:mx}). It is given by
%%%% Missing mass
\begin{equation}
\label{eqn:a}
 M_{x}=\sqrt{(\nu + m_{p} - E_{\pi})^{2} - \abs{\vec{q} - \vec{p_{\pi}}}^{2}},
\end{equation} 
where $m_{p}$ is the proton rest mass; $\nu$ and $\vec{q}$ are energy and momentum of the incoming $\gamma^{*}$, respectively; and $E_{\pi}$ and $\vec{p_{\pi}}$ are energy and momentum of the produced $\pi^{+}$, respectively. Experimental yields were calculated after correcting  for inefficiencies resulting, e.g., from track reconstruction and data aquisition dead times. The HMS tracking efficiency (96-98\%) and pion absorption in the HMS focal plane detectors (2.0\%) were the dominant corrections \cite{fpi2_th,fpi2_hb,fpi2_gh}.

For exclusive meson  electroproduction, the unpolarized  cross section can  be expressed  as a product  of a  virtual photon flux  factor, $\Gamma_{\nu}$,  and a virtual  photon cross section, $\frac{d^{2}\sigma}{d\Omega^{*}_{\pi}}$ \cite{mul}. The reduced  five-fold pion electroproduction cross section is then given by
%%%% Five-fold pion electroproduction cross section
\begin{equation}
\centering
\label{eqn:1.1}
  \frac{d^{5}\sigma}{dE^{\prime}d\Omega_{e^{\prime}}dt d{\phi_{\pi}} } = \Gamma_{\nu} \frac{d^{2}\sigma}{d\Omega^{*}_{\pi}} J(t,\phi_{\pi} \to \Omega^{*}_{\pi}),
\end{equation}
where $E^{\prime}$ and $\Omega_{e^{\prime}}$ are the scattered electron lab energy and solid angle, respectively,  and $\Omega^{*}_{\pi}$ is the pion solid angle in the center-of-mass frame, whereas $J$ is a Jacobian used to transform the cross section in terms of the Mandelstam variable $-t$ and the azimuthal angle $\phi_{\pi}$ between the scattering and reaction planes. The virtual photon flux factor, $\Gamma_{\nu}$, can be expressed as
%%%% Gamma (flux factor)
\begin{equation}
\label{eqn:b}
\Gamma_{\nu} =  \frac{\alpha}{2\pi^{2}}\:\frac{E^{\prime}}{E_{e}}\:\frac{q_{L}}{Q^{2}}\:\frac{1}{1-\varepsilon},
\end{equation} 
where  $\alpha$ is the  fine structure  constant and  $\varepsilon$ is given by 
%%%% Epsilon (longitudinal polarization)
\begin{equation}
\label{eqn:c}
\varepsilon = \Bigg( 1 + \frac{2  \abs{\bf {q}}^{2}}{Q^{2}} \: \tan^{2} \Bigg( \frac{\theta_{e}}{2} \Bigg) \Bigg)^{-1}.
\end{equation}
$q_{L}$ is the equivalent real photon energy, i.e., the  lab energy a  real photon would  require to excite  a target of  mass, $m_{p}$, producing  a system  with invariant mass,  $W$. It is  given in the  Hand convention  by
%%%% Hand convention
\begin{equation}
\label{eqn:d}
q_{L} =  \frac{W^{2} - m^{2}_{p}}{2m_{p}}.
\end{equation}

Furthermore, the information about the hadronic system encoded in $\frac{d^{2}\sigma}{d\Omega^{*}_{\pi}}$ can be expressed as a two-fold unpolarized cross section ($\frac{d^{2}\sigma}{dtd\phi_{\pi}}$) in terms of contributions from longitudinally and transversely polarized photons, and their interference, as follows:
%%%% Rosenbluth Separation
\begin{equation}\label{eqn:1.2}
\begin{split}
  \frac{d^{2}\sigma} {dtd\phi_{\pi}}  & = \frac{1} {2\pi} \bigg( \frac{d\sigma_{T}}{dt} + \varepsilon\frac{d\sigma_{L}}{dt} + \\
                                      &   \quad \sqrt{2\varepsilon(1+\varepsilon)} \frac{d\sigma_{LT}}{dt}\cos{\phi_{\pi}} + \varepsilon\frac{d\sigma_{TT}}{dt}\cos{2\phi_{\pi}} \bigg). 
\end{split}
\end{equation} 
The equation above is often used to separate the different $\sigma_{XX}$ (equivalent to $\frac{d\sigma_{XX}}{dt} $) by means of a Rosenbluth separation. Here, we use it to study the model dependence of the unpolarized cross section, $\frac{d^{2}\sigma} {dtd\phi_{\pi}}$.

%%%% Figure #3 (Missing mass)
%%%%

The determination of the experimental cross section relies on the comparison of the measured experimental yield to the results of Hall C Monte Carlo simulation (SIMC) for the actual experimental set-up, in which a realistic cross section model is implemented.  SIMC traces the reaction products through the spectrometer magnetic fields, incorporates pion decay, energy loss, radiation and multiple scattering effects in the detector elements and other materials in the particle paths, and checks that simulated events cross all required apertures and required detectors before doing a full event reconstruction using realistic detector resolutions.  When  the model input to SIMC describes the dependence of the cross section on all kinematic variables ($W, Q^{2}, -t, \theta_{\pi},$ and $\phi_{\pi}$) correctly (i.e. the ratio of experimental to simulated yield is close to unity within statistical uncertainty), the cross section ($\sigma^{exp}$) for any values of $\overline{W}$ and $\overline{Q^{2}}$ within the acceptance can be determined as
%%%% Yields to Exp. cross section
\begin{equation}
\label{eqn:1.3}
  \bigg( \frac{d^{2}\sigma} {dt \: d\phi_{\pi}}  \bigg)^{exp}_{\overline{W}, \:\overline{Q^{2}}} = \frac{Y_{exp}}{Y_{sim}} \: \bigg( \frac{d^{2}\sigma}  {dt \: d\phi_{\pi}} \bigg)^{model}_{\overline{W}, \:  \overline{Q^{2}}} ,
\end{equation} 
where $Y_{exp}$ is the charge normalized and efficiency corrected experimental yield integrated over  the kinematic acceptance and $Y_{sim}$ is the equivalent simulated yield resulting from the input model cross section. The empirical cross section model provides  the appropriate cross section weighting of the kinematic acceptance and also takes care of bin centering corrections to the experimental cross section \cite{fpi2_hb}. Assuming that the $\phi_{\pi}$-dependence of the cross section is small over the range of $\phi_{\pi}$ in our data, we take just a $t$-dependent function for $\sigma^{model}$. The uncertainty due to this assumption will be discussed later. Given that our data are centered around $\phi_{\pi}=\pi$, the extracted cross section is thus effectively $\frac{d^{2}\sigma}{dt \: d\phi_{\pi}}\rvert_{\phi_{\pi}=\pi}$.

The model cross section for this analysis was  determined using an iterative fitting procedure. The starting pion electroproduction cross section model used in the  simulation is based on a cross section parameterization developed during the F$_{\pi}$-2 L/T separation analysis \cite{q2_scale,fpi2_hb}. The model cross section was  taken as the product of global functions  describing the $W$ and $Q^{2}$ dependences, multiplied by a $t$-dependent  parameterized function for the unpolarized experimental cross sections, i.e., $\sigma = F(W)\cdot H(Q^{2})\cdot G(-t)$. The $W$-dependence is assumed to follow the phase space factor, $F(W) = (W^{2} - m^{2}_{p})^{2}$, based on analyses of the experimental data from Refs. \cite{bebek, brauel}, while the $Q^{2}$-dependence was taken  as $H(Q^{2})=(Q^{2})^{-3.01}$, based on the scaling study of the prior pion electroproduction transverse cross  sections, $\sigma_{T}$ in Ref. \cite{q2_scale}, since $\sigma_{L}$ drops quite rapidly  with increasing $-t$ \cite{q2_scale,cky}. The model  was optimized for $\overline{Q^{2}}=2.4$ GeV$^{2}$ and $\overline{W}=2.0$ GeV to match the $t$-dependence of  the experimental data. The final cross section parameterization for exclusive $\pi^{+}$ electroproduction over the $-t$ range of our data is given by a sum of two $t$-dependent exponential functions as
%%%% Cross section parameterization
\begin{equation}
\label{eqn:1.4}
\begin{split}
  \frac{d^{2}\sigma} {dt\:d\phi_{\pi}} & =  F(W) \: H(Q^{2}) \: \bigg(0.562 \: e^{-5.676 \cdot \abs{t}} \\ 
                                       &    + \quad 0.328 \cdot e^{-1.117 \cdot \abs{t}}\bigg). 
\end{split}
\end{equation}

The coefficients in the above equation are the results of an iterative fit, where the fit of ratios ($Y_{exp}/Y_{sim}$) across all the $-t$ settings was at unity (1.000$\pm$0.013). This model cross section is  valid in the range of $-t$ between 0.272 and 2.127 GeV$^{2}$, and is able to reproduce the cross section results in \cite{clas6} up to $-t=3.0$ GeV$^{2}$ with $\chi^{2}$ per degrees of freedom of 0.94. It is also worthwhile to note that the above equation does not contain a $\phi_{\pi}$ dependence, hence the contributions from $\sigma_{LT}$ and $\sigma_{TT}$ are also present in the cross section results. The significance of $\abs{t}$ coefficients, 5.676 and 1.117 GeV$^{-2}$, is further elaborated in Section \ref{ivc}.

In principle, the extracted unpolarized cross sections are dependent on the model cross section used as input to SIMC; therefore, there is a model dependent systematic uncertainty associated with the extracted $\sigma^{exp}$, partly due to the limited azimuthal coverage of our data set at non-parallel kinematics. In order to study this uncertainty, a $\phi_{\pi}$-dependence as given by Equation \ref{eqn:1.2} was  introduced to the model cross section and the same iterative  procedure was applied. The interference terms, which were based on T. Horn's parameterization \iffalse(\textit{sig\_param3000 from simc; this exact parameterization not found in any of Tanja's literature})\fi that reproduces the F$_{\pi}$-1 (higher $Q^{2}$), F$_{\pi}$-2 and  Brauel \cite{fpi2_hb,brauel} separated cross section data, were assumed to follow
%%%% Interference terms based on prior pi+ data
  \begin{equation}
    \label{eqn:1.5}
    \begin{split}
      \frac{d\sigma_{LT}} {dt} & = \frac{16.533}{(1 + Q^{2})} \cdot e^{(-5.1437 \cdot \abs{t})} \cdot \sin(\theta^{*}), \\
      \frac{d\sigma_{TT}} {dt} & = \frac{178.06}{(1 + Q^{2})} \cdot e^{(-7.1381 \cdot \abs{t})} \cdot \sin^{2}(\theta^{*}). 
    \end{split}
  \end{equation} 
They were calculated using the average kinematic quantities of the data. The $\sigma^{exp}$, determined with the $\phi_{\pi}$-dependence as discussed, were then  compared to  the $\sigma^{exp}$  parameterized by  Equation \ref{eqn:1.4}  to determine  the model dependent  uncertainties for each $-t$ bin. The assigned uncertainty dominates the uncorrelated systematic uncertainty for most $-t$ bins, with an average value of $4.7\%$, while the uncorrelated uncertainties in Table \ref{table:kinematics} reflect the actual value for each bin. Other dominant uncorrelated systematic uncertainties, which affect each $-t$ setting independently, are due to acceptance (0.6\%) and cut dependence (0.5\%), resulting in an average total uncorrelated uncertainty of 4.8\%. 

The correlated systematic uncertainty is predominantly due to radiative corrections \cite{ent01} (2\%), pion absorption (2\%), pion decay (1\%), and acceptance (1\%), resulting in a total correlated uncertainty of 3.5\%. They are listed in Table \ref{table:kinematics}. Both correlated and uncorrelated systematic uncertainties are discussed in more detail in Ref. \cite{fpi2_hb}.  

The statistical uncertainties in the experimental cross  sections are determined by the propagation uncertainties in $Y_{exp}$ and $Y_{sim}$ in Equation \ref{eqn:1.3}.  The statistical uncertainty in $Y_{exp}$ is largely dominated by the uncertainty in the number of  measured real events. However, the uncertainties in the total efficiency, $\varepsilon_{total}$, and in the total  accumulated beam charge, $Q_{total}$, also contribute to the total statistical  uncertainties in measured $Y_{exp}$. The fractional  uncertainty in $Q_{total}$  is 0.5\% for  all the $-t$  settings, while the relative  uncertainty in $\varepsilon_{total}$  is less than  2\%. The statistical uncertainties in both the yield ratio ($R$) and the experimental unseparated cross sections ($\sigma^{exp}$) range from 2\%  to 4\% for all the settings except at the highest $-t$; for that setting, the statistical  uncertainty is close to 6\%. The statistical uncertainty is added in quadrature with the uncorrelated systematic uncertainty to give total random uncertainty. More details on unpolarized cross section determination and uncertainties can be found in Ref. \cite{dada}.

%%%% IV. RESULTS
\section{Results} 
\label{iv}

The unpolarized experimental cross sections, $\frac{d^{2}\sigma}{dtd\phi_{\pi}}$, listed in  Table \ref{table:kinematics} have been extracted with the help of the SIMC, using the relation  in Equation \ref{eqn:1.3} with the model cross section given by Equation  \ref{eqn:1.4}. The two uncertainties for the cross sections in the table are  the combination of statistical and uncorrelated  systematic uncertainties added in quadrature, and  the correlated (scale) uncertainty, respectively.

%%%% Figure #4. (sigL-sigT ratio + cross section result)
\begin{figure}
\centering
\includegraphics[width=\linewidth]{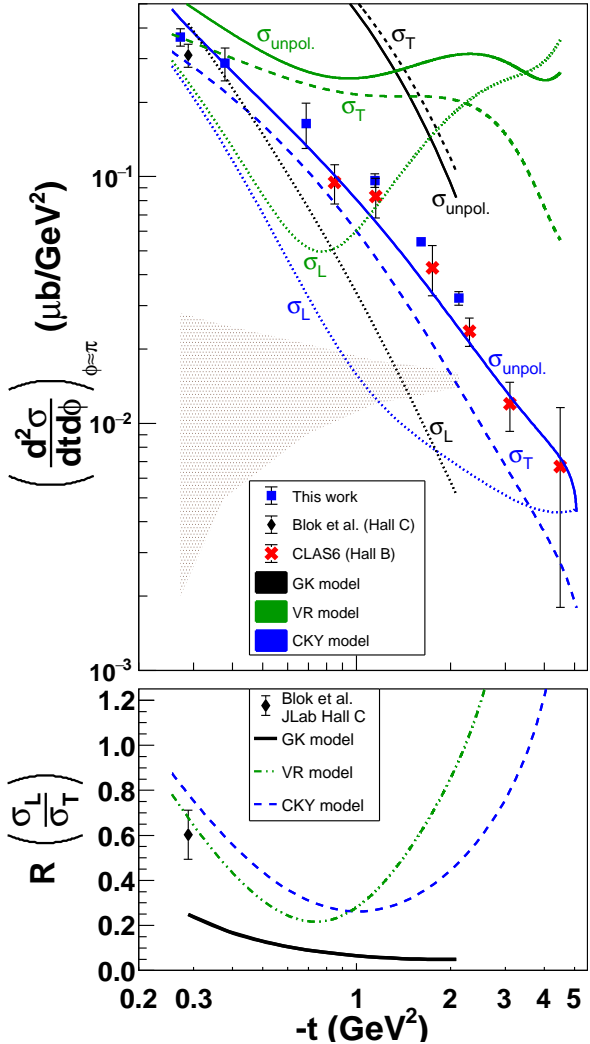}
  \caption{(Color Online) 
{\bf (top panel)} Unpolarized cross sections
($\frac{d^{2}\sigma}{dtd\phi_{\pi}}$) vs $-t$, as the blue squares with error
bars representing total random errors (statistical and random systematic added
in quadrature) of this work. The systematic uncertainties are represented with
shaded band. The data point represented with black diamond at lower $-t$ value
is Blok \textit{et al.}  \cite{fpi2_hb} unpolarized cross section result, while
the red crosses are the kinematically corrected results from JLab Hall B
\cite{clas6}. The results from the Regge amplitude based VR \cite{vr}, CKY
\cite{cky} and GPD-based GK \cite{gk1,gk2} models are also shown in green, blue
and black solid lines, respectively.  Additionally shown are the model
longitudinal (transverse) contributions to the unpolarized cross sections by
dotted (dashed) lines.\\
{\bf (bottom panel)} The ratio ($R = \frac{\sigma_L}{\sigma_T}$) vs $-t$.  The
only L/T separated existing data, from F$_{\pi}$-2 Hall C \cite{fpi2_hb}, are
represented by the black diamond.  The green dot-dashed, blue dashed and black
solid lines represent the VR, CKY and GK models, respectively.  The model
predictions diverge considerably in the region $-t>$ 1 GeV$^2$.
\label{fig:xsc}
}
\end{figure}
%%%%

%%%% 1. COMPARISON WITH PRIOR DATA
\subsection{Comparison with prior data}
\label{iva}

The unpolarized cross section results are compared to two of the prior
exclusive pion electroproduction experiments performed at JLab: F$_{\pi}$-2 L/T
separated (Hall C) \cite{fpi2_th,fpi2_hb} and CLAS (Hall B) data at high $-t$
\cite{clas6}. Both of these measurements were performed with a complete
azimuthal ($\phi_{\pi}$) coverage, which was integrated over in order to
calculate the unpolarized cross sections.

Exclusive $\pi^{+}$ electroproduction above the resonance region was studied
using the CEBAF Large Acceptance Spectrometer (CLAS) in Hall B of JLab in 2001
by scattering a 6 GeV continuous electron beam off a LH$_{2}$ target. The
unpolarized cross sections were measured for the region: $0.16 < x_{B} <
0.58$, $1.6 < Q^{2} < 4.5$ GeV$^{2}$ and $0.1 < -t < 5.3$ GeV$^{2}$ with
complete azimuthal coverage \cite{clas6}. For comparison with the present cross
section data, we consider only those CLAS kinematics that match closely.  These 
span the $-t$ range from 0.85 to 4.50 GeV$^{2}$ at $Q^{2}$ = 2.65 GeV$^{2}$,
$W$ = 2.1 GeV, and $\varepsilon$ and $x_{B}$ of 0.56 and 0.37, respectively. 
As shown in Figure \ref{fig:xsc}, the CLAS data agree quite well with the cross
sections determined with this analysis within the uncertainties.

Even with the limited $\phi_{\pi}$ coverage of our dataset, our cross section,
which was extracted using the ratio method with a $\phi_{\pi}$-independent
model cross section, agrees well with the published cross sections results from
CLAS (Hall B), which follows a different method involving two separate event
generators with the CLAS GEANT3-based Monte Carlo Package, GSIM
\cite{clas6}. The agreement between the two datasets suggests that the 
interference terms contribution is small in this kinematic regime. 
Moreover, the finer binning of the present analysis resulted in smaller 
cross section uncertainties, an improvement compared to the earlier results
from CLAS (Hall B), which were obtained
by coarsely binning their kinematic coverage in $Q^{2}$ and $x_{B}$
\cite{clas6}.

The second pion form factor experiment, F$_{\pi}$-2, was carried out in 2003
with the aim to increase the $Q^{2}$ range of pion form factor from 1.6 to 2.5
GeV$^{2}$. In order to extract the form factor, the separated cross sections
(L, T, LT, TT) were determined using the unpolarized cross sections at two
different $\varepsilon$ values. Thus, in addition to unpolarized cross
sections, we also have access to $\sigma_{L}$ and $\sigma_{T}$ for these data,
where the ratio ($R = \frac{\sigma_L}{\sigma_T}$) between the two is $0.603 \pm
0.117$ at $-t=0.288$ GeV$^{2}$, as shown in the bottom panel of Figure
\ref{fig:xsc}. The unpolarized cross section for $\phi_{\pi}=\pi$ at
$\varepsilon=0.54$ and $-t=0.288$ GeV$^{2}$ is calculated from F$_{\pi}$-2 L/T
separated data using Equation \ref{eqn:1.2}, and compared to the one from this
work. They agree within uncertainties as shown in the top panel of Figure
\ref{fig:xsc}. The F$_{\pi}$-2 L/T separated data compared were at average
values: $\overline{Q^{2}}=2.54$ GeV$^{2}$, $\overline{W}=2.18$ GeV
\cite{fpi2_th,fpi2_hb}. Note, all the data sets were scaled to $Q^{2}$ and $W$
of 2.4 GeV$^{2}$ and 2.0 GeV, respectively, using the $Q^{2}$ and $W$
dependences discussed earlier.

%%%% 2. COMPARISON WITH THE THEORETICAL MODELS
\subsection{Comparison with the theoretical models}
\label{ivb}

In addition to comparing the results from this analysis to previous data, they
were also compared with different available theoretical models.  Historically,
the Regge model by M. Vanderhaeghen, M. Guidal and J.-M. Laget (VGL) \cite{vgl}
is able to provide an accurate description of low $-t$ $\sigma_L$ data in this
$Q^2$, $W$ range, but underestimates $\sigma_T$ by a large factor
\cite{fpi2_th,fpi2_gh}.  Two of the models chosen for comparison were made
after the F$_\pi$-2 L/T separated data were published, with the explicit aim to
provide a much better description of $\sigma_T$ while not destroying the good
description of $\sigma_L$.

The first model used for comparison is the Regge-based model of T. Vrancx and
J. Ryckebusch (the so-called ``VR'' model, where the cutoff masses are fixed by
the authors) \cite{vr}.  This is an extension of the VGL Regge model by the
addition of a hard deep-inelastic scattering (DIS) of virtual photons off
nucleons.  The DIS process dominates the transverse response at moderate and
high $Q^2$, providing a better description of $\sigma_T$.  In
Fig. \ref{fig:xsc} (top panel), the extracted cross section results (including
both F$_\pi$-2 L/T separated and CLAS data sets) are compared with those of the
VR model (solid green line).  The VR model provides a good description of the
low $-t$ region, but fails to describe the unpolarized cross section results
beyond $-t$ of 0.9 GeV$^{2}$.

The second model used for comparison is also Regge-based, by T. K. Choi,
K. J. Kong, and B. G. Yu, the so-called ``CKY'' model \cite{cky}, which uses
$\pi + \rho$ Regge pole exchanges.  In comparison to the VR model, it uses an
alternate set of possible parameters for the pion and the proton charge form
factors to fit the cross sections.  The CKY model uses the cutoff masses for
proton ($\Lambda_{1}$), $\pi$ ($\Lambda_{\pi}$), and $\rho$ ($\Lambda_{\rho}$)
trajectories as three free parameters.  For comparison with the experimental
results, the CKY model cross sections were calculated using the default values
of the cutoff parameters: $\Lambda_{1}=1.55$ GeV, $\Lambda_{\pi}=0.65$ GeV, and
$\Lambda_{\rho}=0.78$ GeV \cite{cky}.  In contrast to the VR model, there is
generally a good agreement between the CKY model (solid blue line) and
unseparated data up to to $-t=4.5$ GeV$^{2}$.  The successful description of
the data from the present work and from CLAS by the CKY model indicates that
the relevant degrees of freedom for our kinematics are hadronic.  The
$\sigma_{L}$ and $\sigma_{T}$ for both VR and CKY models are included in the top
panel of Fig.~\ref{fig:xsc} as dotted and dashed lines, respectively.
The description of the measured $\sigma_L$ is comparable to that of the VGL
model. Although a more detailed comparison of the CKY model and L/T separated
data is needed before a definitive statement can be made, the present level of
agreement promises that the CKY Regge model might be a valid tool for the
extraction of the pion form factor from future electroproduction measurements.

The predictions for $R=\frac{\sigma_{L}}{\sigma_{T}}$ from both models are
shown in the bottom panel of Fig.~\ref{fig:xsc}.  Given they were made after
the publication of the F$_\pi$-2 L/T separated data, it is unsurprising that
$R$ for both models agree with the Hall C data point at $-t$=0.288 GeV$^{2}$
within the uncertainties.  However, as $-t$ increases, $R$, for CKY model,
decreases roughly up to $-t$ of $\sim$ 0.9 GeV$^{2}$ and rises up (with $R >
1.0$) beyond $-t$ of $\sim$ 3.5 GeV$^{2}$. A similar trend is also seen with
the VR model, but $R$ rises much more sharply for VR compared to the CKY model.
%In the absence of separated experimental data in the higher $-t$ region, a
%definite conclusion for $R$ at higher $-t$ cannot be made here.
The large variation in predicted L/T ratios from both models at high $-t$
indicates that this region is very poorly understood.  Much more model
development is needed, and this clearly demonstrates the need for L/T separated
data over a wide $-t$ range.

The experimental results are also compared to the so-called ``GK'' model, a
GPD-based model developed by S.V. Goloskokov and P. Kroll \cite{gk1,gk2}. The
model was developed to study the small $-t$ region at small values of skewness
parameter ($\xi$), where $\xi$, in the light-cone frame, is related to $x_{B}$
by $\xi=\frac{x_{B}}{2-x_{B}}$ and was optimized for exclusive $\pi^{+}$
electroproduction data from HERMES. Thus, the GPDs from the GK model had to be
extrapolated to the higher $-t$ region (up to $-t$ of $\sim$ 2.1 GeV$^{2}$),
which is the kinematic region of our data set. The unpolarized cross sections
were then compared to the model (solid black line) in the top panel of
Fig. \ref{fig:xsc}. Additionally, the longitudinal and transverse contributions
to the unpolarized cross sections are also shown in the figure, with black
dotted and dashed lines and the ratio, $R$, is quite small when compared with
those of the F$_{\pi}$-2 L/T separated data and CKY model, as shown in
Fig. \ref{fig:xsc} (bottom).

It can be clearly seen in the figure that the agreement between our cross
section results and the GK model is quite poor, which can be attributed to the
``handbag'' diagram (and thus, factorization) not being applicable in our
kinematic region. Moreover, the model only makes use of the HERMES exclusive
$\pi^{+}$ data to determine the relevant GPD (helicity-flip, $H_{T}$) and
neglects the sea quark contributions \cite{gk1}. Thus, a revised model making
use of all available $\pi^{+}$ electroproduction data is needed for a better
comparison \cite{kroll}.

%%%% 3. -t SLOPE: A HIGH ENERGY PHYSICS (hep) APPROACH
\subsection{$-t$ slope: A High Energy Physics (HEP) approach}
\label{ivc}

A standard technique used in HEP is the extraction of the exponential slope of the $t$-dependence of the unpolarized cross section, to determine the effective interaction radius a given deep exclusive meson electroproduction reaction is probing \cite{tslope1}. This is done by fitting the unpolarized cross sections with a function of the following form \cite{tslope2}, 
%%%% t-slope
\begin{equation}
  \label{eqn:4.1}
  \frac{d^{2}\sigma} {dtd\phi} = A \cdot e^{- b \cdot \abs{t}},
\end{equation}
where $A$ and $b$ are free parameters. The parameter, $b$, in the above equation can be rigorously linked to the interaction radius for $\gamma^{*}$-$p$ interaction using the equation, 
%%%% Interaction radius
\begin{equation}
	\label{eqn:4.2}
	r_{int} = \sqrt{\abs{b}} \:\:  \hbar c,
\end{equation} 
where $\hbar c$ = 0.197 GeV$\cdot$fm and $r_{int}$ represents the interaction radius. In prior HEP studies \cite{tslope1,tslope2}, this corresponds to the transverse extension of sea quarks and gluons in the proton. We make no such claim in this work, since our data are far from the region where pQCD is expected to be applied. Nonetheless, it provides an interesting insight into the changing character of the reaction in our kinematic regime. 

%%%% Figure #5 (tslope results)
\begin{figure}
\centering
\includegraphics[width=\linewidth]{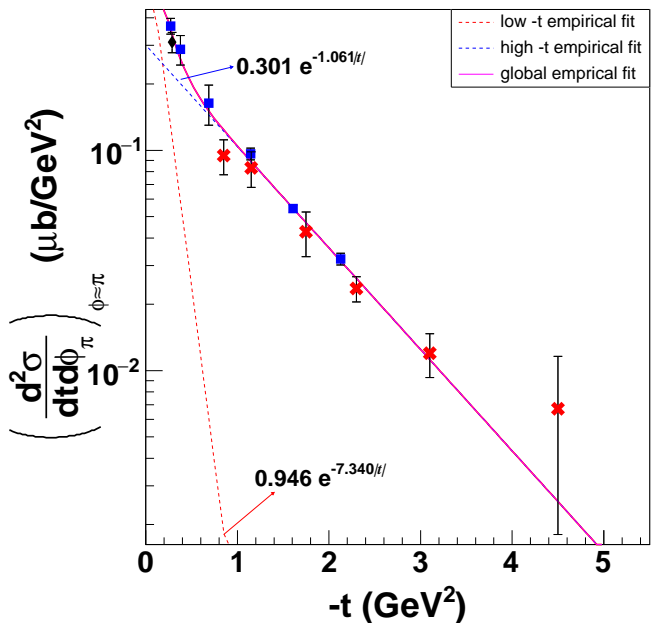}
\caption{(Color Online) The experimental results from Figure \ref{fig:xsc} are shown in log plot. The solid magenta curve, which is the sum of two dotted curves (red and blue) with equation labeled in the plot, is the parameterization of $t$-dependence of combined Halls B and C unpolarized cross section results. The lower $-t$ region is described predominantly by the red dotted curve with steeper $-t$ slope, while the higher $-t$ region is well-characterized by shallower $-t$ slope of the blue curve. See text for details.}
\label{fig:tslope}
\end{figure}
%%%%

%%%% Table II (Impact parameters + interaction radii)
\begin{table}[t!] 
\centering	
\begin{tabular}{lc|cc}
\multirow{3}{1.5cm}{$-t$ Region}& \multirow{2}{2.0cm}{$-t$ Range}& \multicolumn{2}{|c} {\bf{Halls B + C}} \\
           & 		  & $\abs{b}$          & $r_{int}$ \\
	   & (GeV$^{2}$)   & (GeV$^{-2}$)       & (fm) \\ \hline
Low $-t$   & 0 $<-t<$ 0.9 & 7.340 $\pm$ 4.845  & 0.534 $\pm$ 0.176 \\ 
High $-t$  & $-t>$ 0.9	  & 1.061 $\pm$ 0.070  & 0.203 $\pm$ 0.007 \\ 
\end{tabular}
\caption[The exponential slope  of the $t$-dependence of cross  sections and the interaction radii]{The fitted  results for the exponential slopes ($\abs{b}$) of the $t$-dependence of unpolarized $\pi^{+}$ electroproduction cross sections and the interaction radii, along with their respective uncertainties. See text for more details.} 
\label{table:tslope}
\end{table}
%%%%

In the same vein, the $-t$ slope of the unpolarized cross section of the global JLab data is determined by parameterizing the $t$-dependence of cross sections. In Figure \ref{fig:tslope}, the solid magenta curve, which is the sum of two dotted red ($-t$$<$0.9 GeV$^{2}$) and blue ($0.9<-t<5.0$ GeV$^{2}$) curves, represents the parameterization obtained by performing an error-weighted four parameter simultaneous fit to the combined cross section results with an equation of a form: $A \cdot e^{-b_{1} \abs{t}} + B \cdot e^{-b_{2} \abs{t}}$ and the $-t$ slopes are given by parameters $b_{1}$ and $b_{2}$. The corresponding interaction radius, $r_{int}$, was calculated for both $-t$ slopes using Equation \ref{eqn:4.2} and the results are tabulated in Table \ref{table:tslope}. The slope values presented in the table are different than the ones in Equation \ref{eqn:1.4} because here we use all the available data from Halls B and C for the fit. $-t $ of  $\sim$0.9 GeV$^{2}$ was chosen as a transition point between low and high $-t$ regions as the blue curve (corresponding to smaller $r_{int}$ , i.e, harder processes) in Figure \ref{fig:xsc} starts dominating while the contribution from the red curve (corresponding to soft processes) become negligible beyond $-t$ of $\sim$0.9 GeV$^{2}$.

In the low $-t$ region, $r_{int}$ was found to be 0.534$\pm$0.176 fm, which is consistent with the accepted $\pi^{+}$ charge radius of 0.672 fm \cite{pdg}. This corresponds to forward angle meson exchange in the $t$-channel, representing the non-perturbative soft QCD process. In contrast to $r_{int}$ of 0.534 fm in the low $-t$ region, $r_{int}$ is 0.203$\pm$0.007 fm in the high $-t$ region, indicating that the interaction is much harder. In the high $-t$ region, the virtual photon, $\gamma^{*}$, couples directly to parton structure, which is smaller than the radius of electroproduced meson, and hard QCD processes are more important. Ref. \cite{petro} found a very similar interaction radius by analyzing the moments of the response functions of the nucleon from $Q^{2} \approx$ 0.1 to 2.0 GeV$^{2}$, which were interpreted as an effective constituent quark radius. This also is further confirmation of the change in $t$-slopes between low and high $-t$ previously observed in both pion electroproduction \cite{laget} and photoproduction \cite{guidal}. However, the change in $t$-slope occurs at a lower value of $-t$ ($\sim$0.9 GeV$^{2}$) in our work, compared to Refs. \cite{laget,guidal}, where the change to a harder process happens around $-t$ of $2.0-2.5$ GeV$^{2}$.   

%%%% V. CONCLUSION
\section{Conclusion}
\label{v}

The experiment measured exclusive $\pi^{+}$ electroproduction from LH$_{2}$
target over a wide $-t$ range (0.272 to 2.127 GeV$^{2}$) for
$Q^{2}=2.50$GeV$^{2}$ at $W=2.00$ GeV. Assuming small contributions from the
interference terms \cite{clas6,kijun}, two-fold unseparated cross sections were
determined at an average azimuthal angle, $\phi_{\pi}=\pi$, and an average
4.7\% uncertainty due to the incomplete azimuthal coverage assigned. The
experimental results agreed well with prior work from JLab Halls B and C
despite the limited azimuthal coverage of our data, indicating small
contribution from the interference terms in this kinematic region.

The results were also compared to three theoretical models: two using hadronic
approach (VR, CKY), and the other, a partonic approach (GK).  The agreement
between experimental results and the GPD-based GK model is quite poor.  While
it is likely our data have not yet reached the factorization regime, a model
better optimized for high $-t$ kinematics is needed for a better comparison.
The agreement between the experimental results and the CKY model is good within
uncertainties, confirming that the relevant degrees of freedom for our
kinematics are hadronic.  However, the large discrepancy in predicted
$R=\frac{\sigma_{L}}{\sigma_{T}}$ ratios by the VR and CKY models indicate that
even within this picture the high $-t$ region is poorly understood.  An
improved understanding requires L/T separated data over a broad $-t$ range.
The present results validate the method and thus unlock the path to
future high $-t$ L/T separated data.

Following a standard HEP approach, the $-t$ slopes of available JLab data were also calculated in this work, which were used to determine the relevant $r_{int}$ for the $\gamma^{*}$-$p$ interaction. A clear change in $-t$ slope (and $r_{int}$) was observed for the data, indicating the altering nature of the reaction in our kinematic regime. $^{1}$H$(e,e'\pi^+)n$ data up to $-t=0.55$ GeV$^{2}$ with complete $\phi$-coverage will be available in the JLab 12 GeV-era at higher values of $Q^{2}$ and $W$ \cite{fpi3_01,fpi3_02}. The separated $\sigma_{L}$ and $\sigma_{T}$ from those data can be used to further elucidate the soft-hard QCD transition in exclusive charged pion electroproduction.   

%%%% VI. ACKNOWLEDGEMENTS
\section{Acknowledgments} 
\label{vi}

This work was supported by the Natural Sciences and Engineering Research Council of Canada (NSERC) SAPIN-2016-00031 and in part, by the U.S. Department of Energy. We acknowledge additional research grants from the U.S. National Science Foundation, NATO, and FOM (Netherlands). We thank Dr. K. Park for sharing his results from JLab Hall B and Dr. C. Weiss for stimulating discussions on the topic. We also thank Drs. Choi, Kong and Yu for their Regge amplitude based CKY model, as well as Drs. Goloskokov and Kroll of the GPD based GK model for sharing their model calculations and many informative discussions.

%===========================

\end{document}